\documentclass[10pt, conference]{IEEEtran}

\makeatletter
\long\def\@makecaption#1#2{\ifx\@captype\@IEEEtablestring%
\footnotesize\begin{center}{\normalfont\footnotesize #1}\\
{\normalfont\footnotesize\scshape #2}\end{center}%
\@IEEEtablecaptionsepspace
\else
\@IEEEfigurecaptionsepspace
\setbox\@tempboxa\hbox{\normalfont\footnotesize {#1.}~~ #2}%
\ifdim \wd\@tempboxa >\hsize%
\setbox\@tempboxa\hbox{\normalfont\footnotesize {#1.}~~ }%
\parbox[t]{\hsize}{\normalfont\footnotesize \noindent\unhbox\@tempboxa#2}%
\else
\hbox to\hsize{\normalfont\footnotesize\hfil\box\@tempboxa\hfil}\fi\fi}
\makeatother

\usepackage{graphicx}
\graphicspath{{figs/}}
\usepackage{subfigure}
\usepackage{cite}
\usepackage{amsthm}
\usepackage{amssymb}
\usepackage{amsmath}
\usepackage{verbatim}
\usepackage{cite}
\usepackage{graphicx}
\usepackage{url}
\usepackage{setspace}

\usepackage[noend]{algpseudocode}
\usepackage{algorithm}

\newcommand{\etal}{\textit{et~al.}}
\newcommand{\eg}{\textit{e.g.},}
\newcommand{\ie}{\textit{i.e.},}

\title{Optimal Caching and Routing in Hybrid Networks}

\author{Mostafa Dehghan$^{\dag}$, Anand Seetharam$^{\dag}$, Ting He$^*$, Theodoros Salonidis$^*$, Jim Kurose$^{\dag}$, and Don Towsley$^{\dag}$\\

$^{\dag}$School of Computer Science, University of Massachusetts Amherst, $^*$IBM T.J. Watson Research Center  \\
 {\tt \{mdehghan, anand, kurose, towsley\}@cs.umass.edu}, {\tt \{the, tsaloni\}@us.ibm.com}
}

\begin{document}
\maketitle

\begin{abstract}
Hybrid networks consisting of MANET nodes and cellular infrastructure have been recently proposed to improve the performance of military networks. Prior work has demonstrated the benefits of in-network content caching in a wired, Internet context. We investigate the problem of developing optimal routing and caching policies in a hybrid network supporting in-network caching with the goal of minimizing  overall content-access delay. Here, needed content may always be accessed at a back-end server via the cellular infrastructure; alternatively, content may also be accessed via cache-equipped ``cluster" nodes within the MANET. To access content, MANET nodes must thus decide whether to route to in-MANET cluster nodes or to back-end servers via the cellular infrastructure; the in-MANET cluster nodes must additionally decide which content to cache. We model the cellular path as either {\it i)} a congestion-insensitive fixed-delay path or {\it ii)} a congestion-sensitive path modeled as an M/M/1 queue. We demonstrate that under the assumption of stationary, independent requests, it is optimal to adopt static caching (\ie\ to keep a cache's content fixed over time) based on content popularity. We also show that it is optimal to route to in-MANET caches for content cached there, but to route requests for remaining content via the cellular infrastructure for the congestion-insensitive case and to split traffic between the in-MANET caches and cellular infrastructure for the congestion-sensitive case. We develop a simple distributed algorithm for the joint routing/caching problem and demonstrate its efficacy via simulation.
\end{abstract}

\section{Introduction}
\label{sec:introduction}
Future military operations envision processor- and data-intensive real-time applications, including multimedia analytics, situational awareness, location tracking, intrusion detection, and context sharing  to aid data-to-decision at the tactical edge. These services  provide situational awareness in the field and information relevant for mission-critical decisions. 
Increasingly, these applications will operate in hybrid networks consisting of field-deployed tactical MANET nodes and  cellular infrastructure. MANET nodes, equipped with heterogeneous processing and storage capabilities, can communicate among themselves as well as with back-end servers, accessible via the cellular base station. We consider a hybrid network that  supports {\it in-network caching\/}  --  content may always be accessed at a back-end server via  the cellular infrastructure but may also be cached at cluster nodes within the MANET.  Although prior work has demonstrated the benefits of in-network content caching in a wired, Internet (\eg\ CDN) context~\cite{Sitaraman10}, there is limited research understanding the challenges and potential benefits introduced by  caching in hybrid networks~\cite{Golrezaei12, Poularakis13, Westphal13}.

To illustrate the inter-related routing and caching challenges in such a hybrid network, let us consider a data-streaming object/face recognition scenario.  Here, field-deployed MANET nodes continuously capture images of their surroundings and  generate a stream of image-based requests (\eg\ containing features extracted from the image, a timestamp,  physical location, queries), which must then be processed in conjunction with archived images for identification. For example, a  node might identify an object in an image, and want additional information  such as the object's `type' or additional attributes. 
A MANET node must  decide whether to route such an object-identification request to the back-end servers via the cellular infrastructure or to cluster nodes equipped with a cache (Figure \ref{fig:system}). An `object identification module' (executing at both the back-end server(s) accessible via cellular infrastructure and at the in-MANET cache) responds to these requests. At the in-MANET cache, if needed content is available in the cache,  the module can immediately return a reply.  Otherwise, additional needed content must be downloaded (incurring additional delay); the computation can then be performed and  replies can then be returned. Additionally, the cache must decide whether or not to store the downloaded content. 

 {\it The  fundamental question we address in this paper is the following -- how should  nodes route their requests between the cellular infrastructure and the in-MANET cache, and what in-MANET caching policy should  be adopted to minimize expected overall network delay?}
We consider two scenarios, modeling the cellular path as either {\it (i)} a congestion-insensitive fixed-delay path or {\it (ii)} a congestion-sensitive path modeled as an M/M/1 queue. Our goal is to develop delay-minimizing caching and routing policies for this joint routing/caching problem.  Our contributions are as follows.

{\it i)} We prove under the assumption of stationary, independent content requests that it is always optimal to adopt static caching (\ie\ to keep the cache's content fixed over time) based on content popularity (Section ~\ref{sec:prob_soln}).  For the case of fixed delays through the cellular  infrastructure, we show that it is always optimal to route to the  cache for the cached content and to route requests for the remaining content via the cellular  infrastructure.  For the case of congestion-sensitive cellular-access delays, we show that while requests for the cached content should still be routed to the cache, the remaining requests should be split between the cellular infrastructure and the cache;  we also determine the optimal split ratio.

{\it ii)} We present a distributed algorithm (Section ~\ref{sec:dist_soln}) for the joint routing/caching problem, and discuss how both the  nodes and the cache must necessarily be involved in solving this problem. In our distributed algorithm, nodes help the cache infer content popularity; the  cache then uses content popularity to determine what content to cache. Nodes, in turn, determine their routing strategy based on the cached content. Our distributed algorithm seamlessly adapts to changes in content popularity.

{\it iii)} We perform extensive simulations (Section ~\ref{sec:sim}) that demonstrate that our distributed algorithm  provides delay performance comparable to that of the optimal centralized solution.



\section{Related Work}
\label{sec:relatedWork} 

Caching of web-based content has been extensively studied \cite{Liu_98, Che01, Sitaraman10}. Recently, there have also been proposals for designing network-wide caching systems such as Information/Content Centric Networks \cite{Jacobson09}.
In~\cite{Li_13}, the authors develop optimal strategies to provision the storage capacity of routers, balancing the tradeoff between network  delay and provisioning cost. However, the design and analysis of network-level joint caching and routing algorithms, particularly in context of wireless networks, remains an open research problem. 

Prior work has determined  efficient caching and routing policies in cache networks (where every node in the network can cache content and also generate requests for content) \cite{Rosensweig_09, Chai_12, Psaras_12, Sourlas13, Sourlas13_2}. In \cite{Chai_12}, the authors investigate whether a universal network-wide  caching policy is beneficial and quantitatively demonstrate that caching only at a subset of nodes based on network centrality can provide improved performance. Probabilistic caching of content based on parameters such as last requested time and content diversity has been adopted in~\cite{Psaras_12}; the authors show via simulations that such a strategy can provide superior performance.
Azimdoost~\etal~\cite{Westphal13} demonstrate that the asymptotic throughput capacity of a network is
significantly increased by adding caching capabilities to the nodes.
In contrast to the above mentioned works, we consider a system where nodes have two paths (one cached and one uncached) and determine optimal caching and routing policies in this setting.

Our work is closest to~\cite{Golrezaei12, Poularakis13}, where the authors consider the content placement and routing problem in a hybrid network consisting of multiple femtocell caches and the cellular infrastructure with the objective of minimizing delay.  Both papers address the congestion-insensitive case where users have homogeneous delays to the caches, show that their respective problems are NP-hard and propose centralized bounded approximate solutions.
In contrast, we consider a single MANET cache and cellular infrastructure, explicitly model different delays between users and the cache, consider congestion-sensitive cellular path delay and propose a practical distributed algorithm for our joint caching and routing problem. Our caching framework builds on work done for web caches by Liu \etal\ where the authors show that highest cache hit rates are obtained for a static caching policy under independent and stationary request processes~\cite{Liu_98}.

\section{Network Model and Problem Statement}
\label{sec:netmodel}

In this section, we abstract the system described in Section~\ref{sec:introduction}.
We consider a system of $N$ users that generate requests for a set of $K$ unique files
$F~=~\{f_1, f_2, \ldots, f_K\}$ of unit size. We use the terms content and file interchangeably. 

A user $i$ generates requests for the files in $F$ according to a Poisson process of aggregate rate $\lambda_i$.
We denote by $q_{ij}$ the probability that user $i$
generates a request for file $j$ (referred to as the \emph{file popularity}).
Note that the popularity of the same file can vary from one user to another.

All files are available at the back-end server and users are connected to the server via a
cellular infrastructure. We refer to the cellular path between the user and the back-end server as the \emph{uncached path}.
Each user can also access an in-MANET cache where the content might be cached.
Let $C$ be the capacity of the cache measured by the maximum number of files it can store ($C<K$). If user $i$
requests file $j$ and it is present in the cache, then request for file $j$ will be served immediately. We refer to
this event as a cache hit. However, if content $j$ is not present in the cache, then the cache forwards the request
to the back-end server, downloads file $j$ from the back-end server and forwards a reply to the user. We refer to
this event as a cache miss, since it was necessary to download content from the back-end server in order to satisfy the request.

Let $d_{i}^h$ and $d_{i}^m$ denote the delays incurred by user $i$ in the event of a cache hit or miss, respectively.
We assume without loss of generality that  $d_{i}^m  > d_{i}^h$, \ie\ cache misses always incur greater delays than cache hits. We model
the cellular path as {\it (i)}~a congestion-insensitive constant delay path or {\it (ii)}~a congestion-sensitive path
modeled as an M/M/1 queue with service rate $\mu$. In the congestion-insensitive case, the delay experienced
by a request sent by user $i$ is denoted $d_i^0$. We also assume $ d_{i}^h < d_i^0 < d_{i}^m$, for all users $i$.
This is a reasonable assumption because if $d^0_i  < d_{i}^h $, then all traffic for user $i$ will be routed through
the uncached path. On the contrary, if  $d^0_i  > d_{i}^m$, then all traffic for user $i$  will be routed through the
cached path.  In the congestion-sensitive case,  the delay experienced by requests depends on the total incoming
traffic arrival rate and the service rate of the queue (we defer discussion of the M/M/1 queue case until Section~\ref{sec:prob_soln:cong_sen}).

In this work we consider a joint caching and routing problem with the goal of minimizing average content access delay over the requests of all users for all files. The solution to this problem requires addressing two closely-related questions {\it 1)}~How should cache contents be managed -- which files should be kept in the cache, and which cache replacement strategy should be used?
and {\it 2)}~How should the users route (\ie\ split their traffic
for the various files between the cached and uncached paths)?

\section{Deficiency of Cache-Agnostic Routing}
\label{sec:deficiency}

Traditional routing protocols are typically designed to optimize the performance
of \emph{current} traffic load.
We define cache-agnostic routing as the request-routing strategy that routes a request to the (cached or uncached) path that yields the minimum delay given
the current cache content.
Consider the case of the congestion-insensitive uncached path.
It is easy to see that for given content in the cache,
cache-agnostic routing will forward the requests for cached content to the cache,
and route the remaining requests to the uncached path.
We call this the \emph{greedy request routing}.

The greedy nature of cache-agnostic routing makes the system heavily depend
on the initial cache content, since the set of requests that will be routed to the cache depends entirely on files that are already in the cache. Hence, in cases when the cache is preloaded
with unpopular files, greedy routing can perform rather poorly. This phenomenon has also
been observed in other cache networks~\cite{Rosensweig_13}.

The deficiency of cache-agnostic routing can be interpreted from a game-theoretic perspective.
Consider each user as a player in a request routing game where users compete for
the limited cache space to minimize their own delays. Assume the user
population is large so that a single user's decision has
negligible impact on the cache state. Then we observe the following:


{\bf Proposition 1.} \emph{Given any cache state, greedy routing achieves a locally optimal solution}.

This statement is directly implied by the definition of greedy routing because given any cache state,
each user $u$ must route according to
the locally optimal solution, which is given by greedy routing. The routing
game does not have a single local minimum; in fact, each different initial
cache state may lead to a different local minimum. In the next section, we will prove
the following proposition.

{\bf Proposition 2.} \emph{Under the optimal cache state, greedy routing achieves a globally optimal solution}.

The above proposition implies that the global optimum can be achieved by a two-part solution consisting
of centralized cache allocation and distributed greedy routing. This will be the basis of our distributed algorithm
described in Section~\ref{sec:dist_soln}.

\section{Optimal Caching and Routing}
\label{sec:prob_soln}
In this section, we  determine the optimal caching and routing strategies
considering that the uncached path is either {\it i)}~congestion-insensitive, or
{\it ii)}~congestion-sensitive. Overall, we show that it is always optimal
to adopt static caching based on a content popularity metric and to route requests
for cached content to the cache. Requests for the remaining files should be
routed to the uncached path for the congestion-insensitive case, and
should be split between the uncached path and the cache for the
congestion-sensitive case, for which we determine the optimal split ratio. 

\subsection{Congestion-insensitive uncached path} 
\label{sec:prob_soln:cong_insen}
We first consider the case that delays on the uncached path, $d_i^0$ ($d_i^0 > d_i^h$) do not depend on traffic rates.
Let $q_j = \displaystyle \sum_{i=1}^{N} \lambda_i q_{ij}(d^0_i - d^h_i)$ be
the weighted popularity of file $j$. We sort the files in decreasing
order of weighted popularity ($q_j$); let $Q$ be the set of $C$ files with the
highest weighted popularity.
We claim that the optimal caching and routing strategy
is to statically cache the files in $Q$, route requests for these files
to the cache, and route the remaining requests to the uncached path. Let $D_O$ be
the average delay under the optimal policy:
\begin{equation*}
D_O = \displaystyle \sum_{i=1}^{N} \lambda_i \Big( \sum_{j \in Q}q_{ij} d^h_i + \sum_{j \not \in Q} q_{ij}d^0_i \Big).
\end{equation*}

Consider any non-anticipative caching policy $H$, and let $p_{ij}$ be
the fraction of traffic from user $i$ for file $j$ that is routed to the cache.
Let $h_j$ be the hit probability for file $j$ under policy $H$, and let $D_H$
denote the average delay achieved by the non-anticipative caching policy $H$.
It is assumed that a file is immediately available at the cache, even though
there is a delay experienced by the user. For the average delay from $H$ we have
\begin{equation*}
D_H = \displaystyle \sum_{i=1}^{N} \sum_{j = 1}^{K}  \lambda_i q_{ij} \Big( p_{ij}(h_j d^h_i + (1 - h_j)d^m_i) + (1-p_{ij})d^0_i \Big),
\end{equation*}
where $ p_{ij}(h_j d^h_i + (1 - h_j)d^m_i)$ and $(1-p_{ij})d^0_i$ are the average delays for requests from user $i$ and file $j$ routed to the cache and the uncached path,
respectively. To demonstrate optimality,
we state the following lemma. The proof is given in the appendix.

{\bf Lemma 1.} \emph{$D_H \ge D_O$}.

Our proof borrows ideas from prior work~\cite{Liu_98}, where the authors prove
that given the traffic access rates for files, statically caching the
$C$ files with the highest popularity will result in the highest hit rate.
Note that a corollary to Lemma~1 is that optimal delay can be achieved
by loading the cache with these $C$ most popular files and having the users greedily route
requests for content.

\subsection{Congestion-sensitive uncached path}
\label{sec:prob_soln:cong_sen}
Let us next consider the case where delays on the uncached path are congestion-senstitive, \ie\ they depend on the request rate on that path. We assume that hit and miss delays are equal among all users, \ie\
$d^h_i= d_h$ and $d^m_j = d_m$, respectively. The uncached path is modeled as an
M/M/1 queue with service rate $\mu$. We assume that the average service time in
the queue lies between the hit and miss delays, \ie\ $d_h < 1/\mu < d_m$. We note
that the average delay through an M/M/1 queue with incoming rate $\lambda$ and
service rate $\mu$ is given by $1/(\mu -\lambda)$ when $\lambda < \mu$.

Let $q_j = \displaystyle \sum_{i=1}^{N} \lambda_i q_{ij}$ denote the overall
popularity of file $j$. We sort the files in decreasing order of overall popularity
($q_j$); let $Q$ be the set of $C$ files with the highest overall popularity.
Since the hit delay is smaller than the miss delay and delay from the uncached
path, a reasoning similar to that in~\ref{sec:prob_soln:cong_insen} again implies that
it is optimal to
have the files in $Q$ in the cache, and route requests for those files to the
cache.

In the remaining, we show that to achieve optimal delay it is necessary to split the
traffic for the uncached content between the cache and the uncached path.
Note that when the uncached path is congestion-sensitive, sending all the traffic
for uncached files could potentially congest that link, resulting in a large delay. Therefore, a portion of the traffic for uncached files should be
directed to the cache, incurring a miss there and the consequently an access delay $d_m$.
It is important to note that the cache content is {\it not} updated in case of a cache miss.
Assume that user $i$ routes a 
fraction $p_{ij}$ of the traffic for the uncached file $j$ to the cache.
The expected delay can be expressed as
\begin{eqnarray*}
D_O&=&  \sum_{i=1}^N\sum_{j \in Q} \lambda_iq_{ij} d_h \nonumber + \sum_{i=1}^N \displaystyle \sum_{j \not \in Q}  \lambda_i q_{ij}p_{ij} d_m \\
&& + \frac{ \displaystyle\sum_{i=1}^N  \sum_{j \not \in Q}  \lambda_i q_{ij} (1 - p_{ij}) } {\mu -  \displaystyle  \sum_{i=1}^N \sum_{j \not \in Q}   \lambda_i q_{ij} (1 - p_{ij})  }
\end{eqnarray*}
The above function is convex and can be differentiated with respect to $p_{ij}$:
\begin{equation*}
\frac{\partial D_O}{d p_{ij}} = \lambda_i q_{ij} d_m -  \frac{\mu  \lambda_i q_{ij}}{(\mu - \displaystyle\sum_{i=1}^N  \sum_{j \not \in Q}  \lambda_i q_{ij} (1 - p_{ij}))^2} 
\end{equation*}
Equating the derivative to zero we get
\begin{equation}
\label{eq:p_sum}
\sum_{i=1}^N \displaystyle \sum_{j \not \in Q} \lambda_iq_{ij} (1-p_{ij}) = \mu - \sqrt{\frac{\mu}{d_m}}
\end{equation}
In the above formulation, the optimal solution depends only
on the amount of traffic being routed through the congestion-sensitive link and
not the type of the file being routed to the uncached path. The optimal
delay is achieved for any values of $p_{ij}$ that satisfy~\eqref{eq:p_sum}, and  different sets of routing probabilities will yield in this (same) optimal delay value. One
such solution occurs when $p_{ij} = p$, and from~\eqref{eq:p_sum} we get
\begin{equation}
\label{eq:p}
p = 1- \frac{\mu - \sqrt{\frac{\mu}{d_m}}}{\displaystyle \sum_{i=1}^N \displaystyle \sum_{j \not \in Q}  \lambda_iq_{ij}}.
\end{equation}

\section{Distributed Algorithm}
\label{sec:dist_soln}
Based on our analysis in the previous section, the global optimum can be reached by a two-step solution consisting of a centralized static cache allocation followed by distributed greedy routing.  The centralized caching solution relies on the existence of a central authority that oversees all the user demands and controls all the caches. This approach may be difficult to apply in a MANET environment where it is desirable to have a distributed solution that only relies on local information. Furthermore, in practice user demand can change over time due to user mobility or changes in file popularity. Therefore, a challenge in solving the joint routing and caching problem in a distributed manner is for the cache to infer the file popularities (from the individual file popularities of users) in order to decide which files to cache; the routing of requests will in turn depend on the cached files. 

In what follows, we present a Distributed Caching and Routing (DCR) algorithm, which aims to emulate the behavior of the optimal joint caching and routing policies and to seamlessly adapt to changes in content popularities.  Due to space limitations and for clarity purposes, we present the algorithm for the case where \emph{the delays are equal across different users}. Our algorithm can be easily generalized to the case of different delays for different users.

The distributed algorithm consists of two \emph{phases}. The
first phase corresponds to the state when the cache has an inaccurate estimate
of the file popularities, during which it will observe a fixed portion of the traffic for
all files. We call this state the ``\emph{caching phase}'' and require the users
to send a specific fraction, $\alpha$, of their traffic to the cache. This allows the cache to estimate the file popularities as well as the aggregate request
rate of the users. We determine the value of $\alpha$ such
that the average delay during the caching phase is minimized.

The second phase, called the ``\emph{routing phase}'', begins after the cache
gathers enough data regarding user traffic. At this point the cache is able
to estimate file popularity and update the cache content. Note that users can
learn whether a content is stored in the cache or not based on the difference
between the hit and miss delays. If users learn that some content is in the cache,
they will always forward their traffic for those content to the cache. However,
for content that is known not to be in the cache, a fraction $p$ (specified
by the cache) of the traffic gets forwarded to the cache, and the remaining is
routed to the uncached path. The reason for splitting the traffic for uncached
content is twofold: 1)~it provides a means for the cache to estimate
the traffic for uncached content when popularities change over time,
and 2)~avoids congesting the uncached path in the congestion-sensitive
case.

Note that in the caching phase the cache can observe a fixed fraction of the
traffic for all files, and can estimate content popularity based on the incoming
traffic. The cache estimates the aggregate arrival rate
$\hat{\lambda}$ by dividing the number of observed arrivals per time unit by $\alpha$.
Note that in the routing phase only a fraction $p$ of the
traffic for the uncached content is observable to the cache. Since this portion
of the traffic mainly corresponds to the misses at the cache, the cache can
estimate file popularities $\hat{q} = \frac{1}{\Lambda} (n_h + n_m / p)$,
where $n_h$ and $n_m$ denote vectors containing the number of observed requests
resulting in hits and misses for different files, and $\Lambda$ is
a normalizing constant. More sophisticated techniques can be used to
estimate popularities (\eg\ see~\cite{Good53}), but we will see in the next section
that this simple approach suffices in our case.

Algorithm~\ref{alg:distributed} summarizes the steps for the distributed caching
and routing described above. Note that at any point in time, the cache can start
the caching or routing phase by broadcasting a message to users.
We discuss next how to select the two parameters $\alpha$ and $p$. 

\begin{algorithm}
\caption{Distributed Caching and Routing (DCR)}
\label{alg:distributed}
\begin{algorithmic}[1]
	\Statex // Caching Phase
		\State Cache broadcasts the ``caching phase'' message with parameter $\alpha$.
		\State Users send requests to the cache with probability $\alpha$, and to the back-end server with probability $1 - \alpha$.
		\State Cache estimates the file popularities and aggregate request rate.
	\Statex // Routing Phase
		\State Cache broadcasts the ``routing phase'' message with parameter $p$.
		\State Based on response times from the cache, users decide whether a file is
		in the cache or not.
		\State For the files that (users think) are in the cache, users send their requests
		to the cache.
		\State For the files not in the cache, users send requests to the cache with
		probability $p$, and to the back-end server with probability $1 - p$.
\end{algorithmic}
\end{algorithm}

\subsection{Congestion-insensitive uncached path}

Let $\bar{d}_c$ be the average delay assuming that all users route their entire traffic to the cache.
During the caching phase, as users route $\alpha$ portion of their traffic to the cache, the average delay ($D$) is given by $D = \alpha \bar{d}_c + (1 - \alpha) d_0$.
It is easy to see that the average delay is minimized for $\alpha = 0$ if
$\bar{d}_c > d_0$, and for $\alpha = 1$, otherwise. Since, for estimation purposes,
we require $\alpha$ to be greater than zero, we choose $\alpha = 0.5$ noting that
it results in a delay no more than a factor two of the optimal.

For the routing phase, it can be easily seen that the optimal delay is achieved
by having $p = 0$. However, this optimality is achieved under the assumption of
static content popularities. In order to make it possible for the cache to estimate
the traffic for uncached content and track changes in content popularity, we fix
this parameter at $p=0.1$ to direct $10\%$ of the traffic for uncached content to
the cache.

\subsection{Congestion-sensitive uncached path}
For the congestion-sensitive case, the average delay ($D$) over the caching phase can
be written as
\[D = \alpha \bar{d}_c + \frac{1 - \alpha}{\mu - (1 - \alpha) \lambda},\]
where $\lambda$ is the aggregate request rate, $\mu$ is the service rate of the
uncached path, and $\bar{d}_c$ is the expected delay from the cache and can be
estimated as
$\hat{\bar{d_c}} = (\hat{\lambda'} d_h + (\hat{\lambda} - \hat{\lambda'}) d_m) / \hat{\lambda}$, where $\hat{\lambda}$ and $\hat{\lambda'}$ are estimates of the aggregate request rate and the request rate for the cached content, respectively. $\hat{\lambda'}$
is estimated similar to $\hat{\lambda}$.
Assuming that $\mu$ is known to the cache, $\alpha$ can be computed as
\[\alpha = \frac{1}{\hat{\lambda}}(\sqrt{\frac{\mu}{\hat{\bar{d_c}}}} - \mu + \hat{\lambda}),\]
to minimize $D$. If the cache does not have any (or accurate) estimates for $\lambda$ or
$\lambda'$, it is desirable to start with a fairly large value for $\alpha$ to prevent
congesting the uncached path.

For the routing phase, the parameter $p$ can be computed based on~\eqref{eq:p}.
However, to prevent the uncached path from getting congested, we always take the value
$p = \max(0.1, p')$ where $p'$ is computed using~\eqref{eq:p}.

\section{Performance Evaluation}
\label{sec:sim}

In this section, we use simulation to  {\it i)} demonstrate the importance of jointly optimizing caching and routing rather than optimizing with respect to just one of these considerations, and {\it ii)} demonstrate the efficacy of DCR by showing that its delay performance is comparable to the optimal algorithm.

\subsection{Centralized Solution}
To show the need for jointly optimizing caching and routing, we consider a caching system with a congestion-insensitive path to the base station, and evaluate the average delay achieved under various caching and routing policies. The policies outlined below are centralized and  we consider policies which optimize over neither, either or both caching and routing.

\noindent{\it LRU:}
We assume that the cache implements the Least-Recently-Used replacement policy. The
routing is simple, with users sending all traffic to the cache (note that the uncached
path is not utilized here). We use this scenario as a baseline case for evaluation of the
distributed algorithm.

\noindent{\it Optimized Caching:} 
We assume that the cache statically caches the most popular files, and that
all requests are routed to the cache. Note that the caching policy here corresponds
to the optimal caching policy, while the routing policy is naive.

\noindent{\it Optimized Routing:}
In this case, the cache replacement policy is LRU. To determine the routing
strategy, users determine the expected delay of requests to the cache assuming
that all traffic is routed to the cache. This expectation is calculated by using the
approximation for determining the hit rate for the different files outlined in~\cite{Che01}.
Users route requests for each file along that path (cached/uncached) that has a
lower expected delay for that file.

\noindent{\it Optimal:}
In this case, the cache statically caches the most popular content. Users
send traffic for the cached content to the cache and for the remaining files
to the uncached path. As shown earlier this policy yields the optimal (minimum) delay.

\begin{figure}[t]
\begin{center}
  	\includegraphics[scale=0.3]{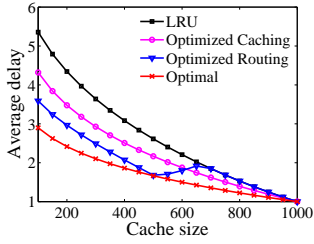}
  \caption{Comparison of the average delay for jointly solved caching and routing problem (Optimal) with partially optimized solutions.}
 \vspace{-0.5cm}
    \label{fig:centralized}
\end{center}
\end{figure}

Figure~\ref{fig:centralized} plots the average delay achieved by the policies explained above for different values of the cache size. We assume that $d_h = 1$, $d_m = 8$ and $d_0= 5$ time units. We consider 5 users generating requests for 1000 files with file popularities having a Zipf distribution with skewness parameter 0.8. We observe from the figure that {\it LRU} performs poorly for small cache sizes, with the performance improving as the cache size increases. {\it Optimized Caching}, which optimizes only caching and not routing, performs  better than {\it LRU}, but is not close to {\it Optimal}.

{\it Optimized Routing} combines a traditional caching policy (LRU) with a greedy routing policy  in which each user determines apriori what files to route to the cache regardless of the actual traffic being routed to the cache by other users. We observe that when the cache size is small, the performance of {\it Optimized Routing} is poor compared to {\it Optimal}. As the cache size increases, the performance of {\it Optimized Routing} improves and it equals the performance of {\it Optimal} for a cache size of 500. The reason for this behavior is that, for small cache sizes
{\it Optimized Routing} results in users sending requests for unpopular and relatively popular files through the uncached path leaving the cache partially empty. This behavior persists until a cache size of 500 when {\it Optimized Routing} requests the most popular files from the cache and provides the same performance as {\it Optimal}. Beyond a cache size of 500, this policy routes requests for a larger number of files (greater than the cache size) to the cache which results in a behavior similar to {\it LRU}. We observe that this zigzag behavior of {\it Optimized Routing} will hold regardless of the simulation parameters; the number 500 is an artifact of the particular parameter values of this simulation.

In comparison to all the other algorithms, the delay obtained from {\it Optimal} is  smaller. Our simulations show that to minimize user delay it is of prime importance to optimize both caching and routing; optimizing one and not the other can result in suboptimal and often unsatisfactory delay performance.

\subsection{Distributed Solution}
Having demonstrated the importance of optimizing joint caching and routing, our next goal is to show that our distributed algorithm DCR achieves delay performance similar to {\it Optimal}.  
We once again consider a group of 5 users, 1000 files, $d_h = 1$ and $d_m = 8$.
The demand for each user follows a Poisson process with aggregate rate $\lambda=1/5$. The initial file popularities follow a Zipf distribution with skewness parameter 0.8.

We simulate changes in file popularities that happen with probability 0.01 at each request arrival. For each change, we select a random user $u$ and request type $f$, and update the popularity $q_{uf}$ as $q_{uf} = \min(1, \max(0, q_{uf} + \Delta))$ where $\Delta$ is uniformly distributed in $[-\Delta_{\max}, \Delta_{\max}]$, and $\Delta_{\max} = q_{uf}$. 
We assume  $d_0 = 5$ and $\mu = 0.5$ in the congestion-insensitive and congestion-sensitive cases, respectively.

To evaluate the performance of the proposed distributed algorithm, we compare the delay performance  of the Distributed Caching and Routing (DCR) algorithm with  two other algorithms -- the  Distributed Caching and Optimized Routing (DCOR) algorithm and {\it Optimal}. DCOR consists of a `caching phase', where the cache estimates the  file popularities and caches the popular files. We assume that users have perfect knowledge of the cache state and route requests greedily based on whether the content is present in the cache or not. {\it Optimal} is similar to the one described in the previous section (\ie\ before every request arrives, the algorithm determines the optimal set of files to be placed in the cache based on the current popularities and then determines the routing based on the cached content). Note that DCOR and {\it Optimal} are not implementable in practice.

\begin{figure}[t]
\begin{center}
  	\includegraphics[scale=0.3]{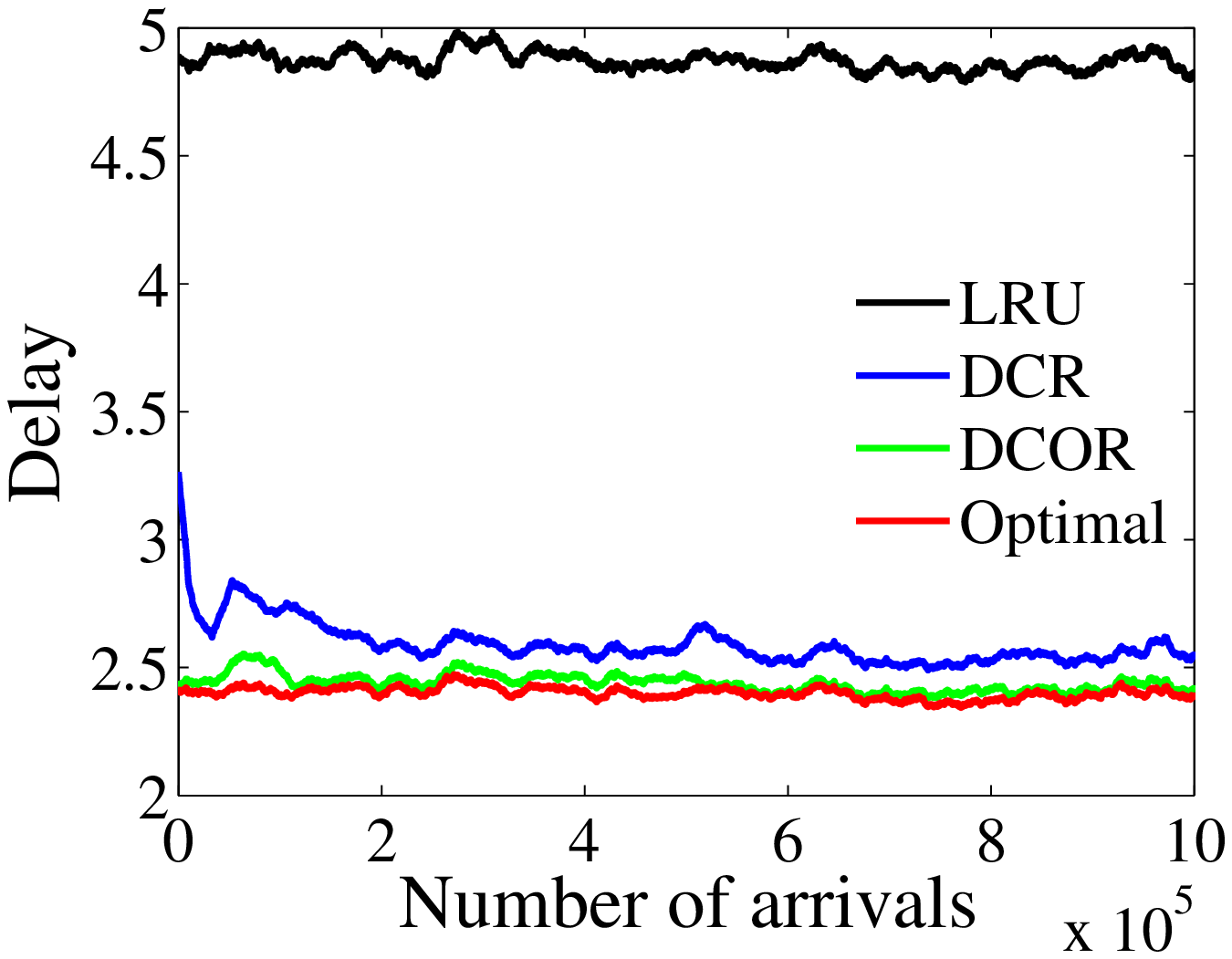}%
	\includegraphics[scale=0.3]{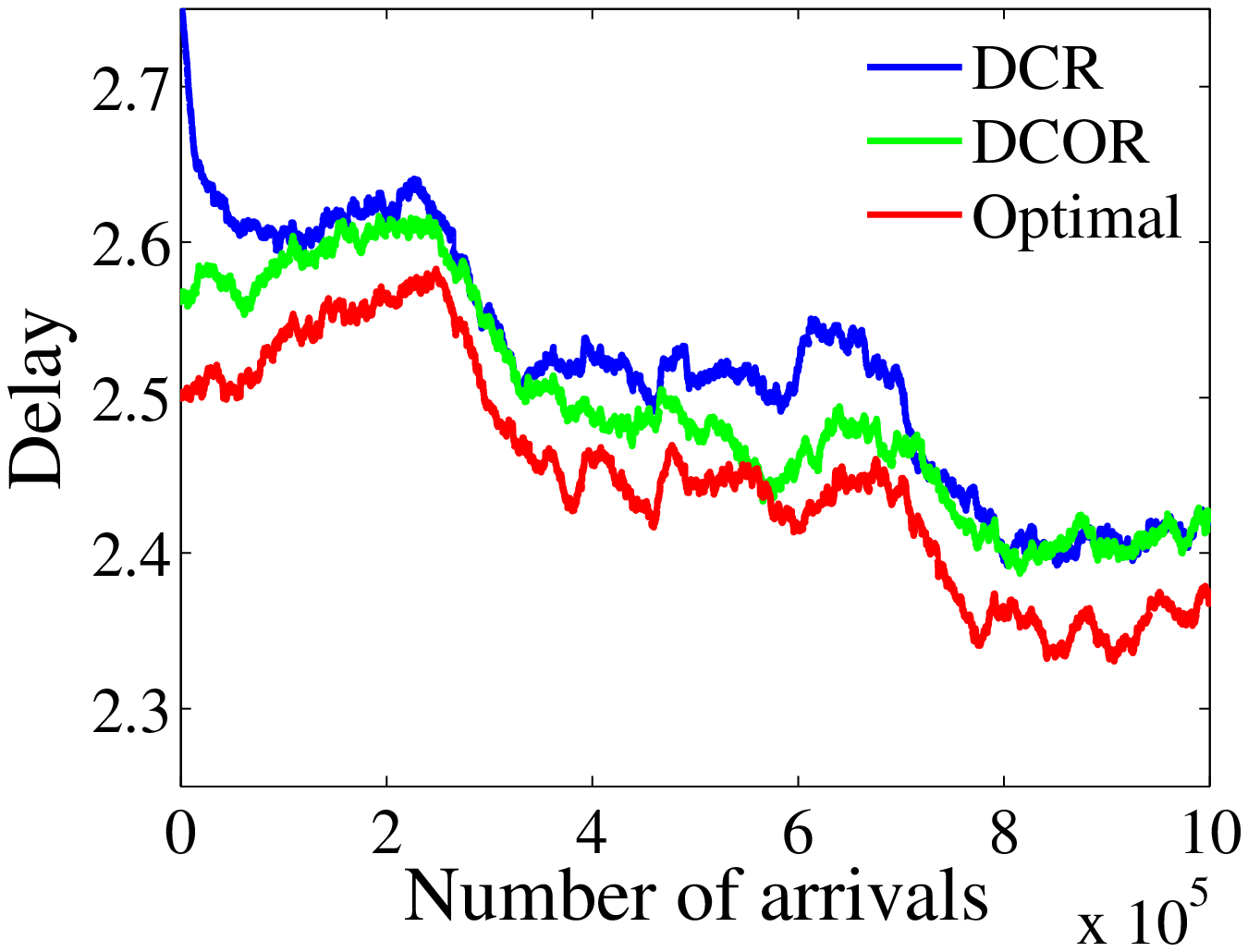}
  \caption{Comparing the average delay obtained by the distributed algorithm with delay achieved by {\it Optimal} and DCOR for congestion-insensitive (left)
  and congestion-sensitive (right) delay models for the uncached path.}
 \vspace{-0.5cm}
    \label{fig:delay}
\end{center}
\end{figure}

Figure~\ref{fig:delay} compares the delay values obtained from the DCR, DCOR and {\it Optimal} algorithms over $10^6$ arrivals for the congestion-insensitive and congestion-sensitive uncached path delay models. For the congestion-insensitive case, we present delay values for \emph{LRU} which is clearly far from {\it Optimal}. We also evaluate \emph{LRU} for the congestion-sensitive case, but we omit it from the figure as its delay performance is further away from {\it Optimal}.

The purpose of exploring the different algorithms is to separately determine the impact of  imperfect caching and routing on the loss in performance. The difference between {\it Optimal} and DCOR depicts the loss in performance as a result of imperfect caching. We observe that  there is negligible difference between DCOR and {\it Optimal} which indicates that even our naive method of estimating popularities performs well in practice. The difference between DCR and DCOR is primarily due to the imperfect knowledge of  the cached content by the users; it can be seen that as the number of arrivals increase, users have a better understanding about the cached content and DCR performs closer to DCOR  and {\it Optimal}. Further we also observe that the relative performance of DCR  is better for the congestion-sensitive case than the  congestion-insensitive case. 

\begin{figure}[]
\begin{center}
  	\includegraphics[scale=0.3]{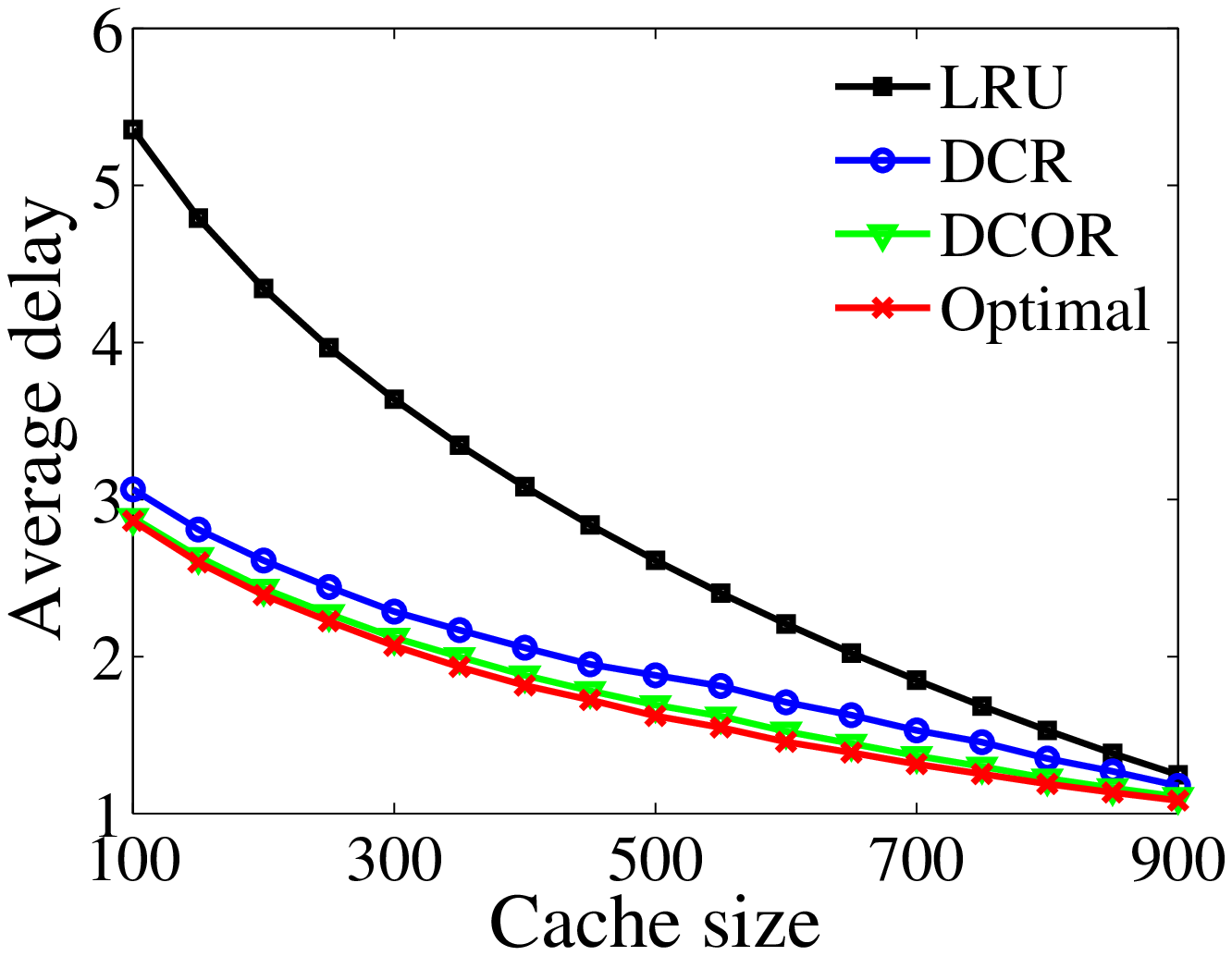}%
	\includegraphics[scale=0.3]{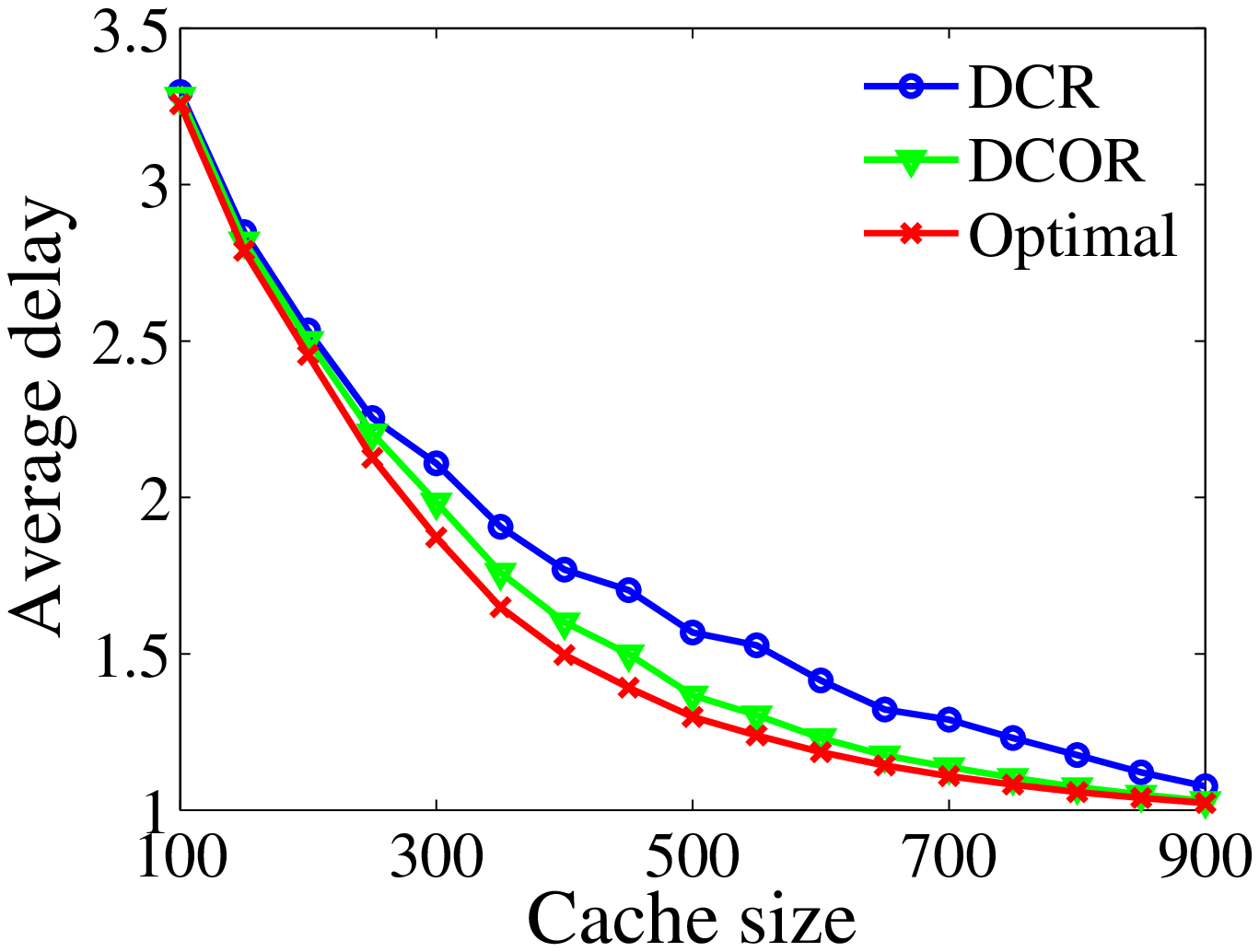}
  \caption{Average delay versus the cache size for congestion-insensitive (left)
  and congestion-sensitive (right) delay models for the uncached path.}
\vspace{-0.5cm}
    \label{fig:delay_cache_size}
\end{center}
\end{figure}

To get a better sense of expected performance of the DCR algorithm, we evaluate the
average delays achieved by DCR and its benchmarks as a function of the cache size.
Figure~\ref{fig:delay_cache_size} shows the results averaged over 10 runs of the simulation
(each run consists of $10^6$ arrivals). As expected, the average delay decreases as the cache
size increases. It can be seen that DCR performs very close to {\it Optimal}.
The reason DCR deviates from the {\it Optimal} for larger cache sizes (compared with the total number of files) is that we restrict the
parameter $p$ to 0.9 as described in Section~\ref{sec:dist_soln}. When the cache size is
large, a larger portion of the traffic can be satisfied from the cache and it is less
likely that the uncached path will be congested. However, in practice we are more
interested in cases where the cache size is small compared to the number of all files.

\section{Conclusion}
\label{sec:conclusion}

In this paper, we studied a joint routing and caching problem in a hybrid network, consisting of MANET nodes and cellular infrastructure. We modeled  the cellular path as either {\it (i)} a congestion-insensitive constant delay path or {\it (ii)} a congestion-sensitive path modeled as an M/M/1 queue and demonstrated  that it is always optimal to adopt static caching based on content popularity.  We also showed that it is always optimal to route for cached content to the cache; for the remaining  content,  requests should be routed to the  cellular path  in the  congestion-insensitive case and should be  split between the cellular infrastructure and the cache  in the  congestion-sensitive case. We also developed a simple distributed algorithm for the above problem and illustrated its performance via simulation.

\bibliographystyle{IEEEtran}
\bibliography{articles1}
\appendix[Proof of Lemma 1]
\label{sec:appendix}
\begin{eqnarray*}
D_h - D_o &=& \displaystyle \sum_{i=1}^{N}\big ( -\sum_{j = 1}^{K} \lambda_i p_{ij}q_{ij}h_j(d^m_i - d^h_i) \\
&&+ \sum_{j = 1}^{K} \lambda_i p_{ij}q_{ij}(d^m_i - d^0_i)  +  \sum_{j \in Q}\lambda_i q_{ij}(d^0_i - d^h_i)  \big)\\
&= &  \displaystyle \sum_{i=1}^{N}  \big (\sum_{j = 1}^{K} \lambda_i p_{ij}q_{ij}(1-h_j)(d^m_i - d^0_i)  \\
&& +   \sum_{j \in Q}\lambda_i  q_{ij}(d^0_i - d^h_i)  - \sum_{j = 1}^{K}\lambda_i  p_{ij}q_{ij}h_j(d^0_i - d^h_i) \big ).
\end{eqnarray*}
\\
Clearly, $\displaystyle \sum_{i=1}^{N} \sum_{j = 1}^{K}  \lambda_ip_{ij}q_{ij}(1-h_j)(d^m_i - d^0_i)  \geq 0$  and we need to prove that 
\[A = \displaystyle \sum_{i=1}^{N}  \sum_{j \in Q} \lambda_i\big (q_{ij}(d^0_i - d^h_i)  - \sum_{j = 1}^{K} p_{ij}q_{ij}h_j(d^0_i - d^h_i)   \big ) \geq 0.\]
We have,
\begin{eqnarray*}
 A &=& \displaystyle \sum_{i=1}^{N}\big ( \sum_{j \in Q} \lambda_i  q_{ij}(d^0_i - d^h_i)  - \sum_{j = 1}^{K}\lambda_i  p_{ij}q_{ij}h_j(d^0_i - d^h_i)  \big ) \\
& \geq & \displaystyle \sum_{i=1}^{N} \big ( \sum_{j \in Q} \lambda_i q_{ij}(d^0_i - d^h_i)  - \sum_{j = 1}^{K}  \lambda_i q_{ij}h_j(d^0_i - d^h_i)  \big ) \\
& = &   \sum_{j \in Q}q_j - \sum_{j = 1}^{K} h_jq_j \\
& \geq & 0 
\end{eqnarray*}
Therefore $D_h \geq D_o$.

\end{document}